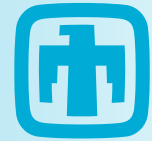

# Synchronic Web Digital Identity

**Speculations on the Art of the Possible**

Thien-Nam Dinh    Justin Li    Mitch Negus    Ken Goss

thidinh@sandia.gov    jdli@sandia.gov    mnegus@sandia.gov    kgoss@sandia.gov

March 12, 2024 (last modified May 21, 2025)

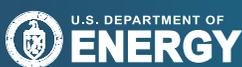
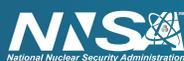

## ABSTRACT

As search, social media, and artificial intelligence continue to reshape collective knowledge, the preservation of trust on the public infosphere has become a defining challenge of our time. Given the breadth and versatility of adversarial threats, the best—and perhaps only—defense is an equally broad and versatile infrastructure for digital identity.

This document discusses the opportunities and implications of building such an infrastructure from the perspective of a national laboratory. The technical foundation for this discussion is the emergence of the Synchronic Web, a Sandia-developed infrastructure for asserting cryptographic provenance at Internet scale. As of the writing of this document, there is ongoing work to develop the underlying technology and apply it to multiple mission-specific domains within Sandia. The primary objective of this document to extend the body of existing work toward the more public-facing domain of digital identity.

Our approach depends on a non-standard, but philosophically defensible notion of identity: digital identity is an unbroken sequence of states in a well-defined digital space. From this foundation, we abstractly describe the infrastructural foundations and applied configurations that we expect to underpin future notions of digital identity.

# CONTENTS

# 1. INTRODUCTION

This section describes the conceptual preliminaries that motivate the remaining sections. We provide this section only to establish context for subsequent technical decisions. Detailed discussion of the sociological accuracy, coherence, or utility of these statements are outside the scope of this document. We first define the two key concepts that underpin the technical design.

- **Digital Identity**: A digital identity is a sequence of semantically linked computational states over time that corresponds to a real-world entity. [1] For example, the sequence of all posts and direct messages involving a specific social media handle, and the sequence of all transactions involving a specific smart contract address, a database containing all historical communications of a small business, are all digital identities. The two salient criteria are (1) that the sequence of states all exist on the same information system, and (2) that the sequence of states is complete. [2].

- **Digital Reputation**: A digital reputation is the confidence value that an digital identity is willing to assign to an unknown statement by another digital identity within some known context. We specify that the statement is unknown to isolate beliefs about identities from beliefs about the content of the statement itself. For example, the confidence that a social media user has in a medical recommendation issued by a government agency is the reputation of the agency to the user in the medical context. In our definition, reputation is a pairwise, rather than global, concept.

In the real world, identity and reputation depend on both clearly objective and potentially subjective information. The focus of our work is to model and reinforce the purely objective aspects. In particular, the two objectively knowable facts about any given statement are: (1) who made the statement and (2) when the statement was made. Together, these two pieces of information determine the *provenance* of the statement. In contrast, none of the techniques in this document attempt to directly address the subjective truth of statements. However, we hypothesize that a strong objective foundation is a necessary step toward optimally reaching social consensus over the subjective components. Finally, much has been written about the dangers that technology poses for society, for instance, about the privacy implications of digital identity or the democratic implications of digital reputation. Normative thoughts about whether society should continue to build these systems is out of scope for this document. Instead, given that these systems already exist, we only seek to contribute improvements to the technical design space.

---

[1] This definition attempts to mirror Derek Parfit's characterization of personal identity ("the self") as a sequence of continuously connected psychic states over time [11]

[2] The notion of identity is related to, but distinct, to the notions of *identifiers* and *presentations*. An identifier is a string that refers to a digital identity at some point in time. A presentation is a subset of the digital identity that is revealed in some digital interaction.

## 2. BACKGROUND

This section describes relevant prior work in two broad categories: digital identity and decentralized web. Our objective is not to provide an academically comprehensive literature review, but to provide an informative sample of the existing landscape.

### 2.1. Digital Identity

This section covers concepts that most security professionals would categorize as "digital identity" (in contrast to our relatively more expansive definition). Relevant work in this field is best understood in the context of the common standards that underpin existing identity and access management technologies. The following are select standards from the three most relevant international standards-making bodies.

- **IETF**: The IETF (Internet Engineering Task force) is the primary standards-development body for the Internet. The relevance to this document is its coverage of lower-level networking protocols. For example, *System for Cross-domain Identity Management* standard specifies the exchange of digital identity information across enterprise information systems [6]. *Remote Attestation System* specifies the process by which one information system assesses the trustworthiness of a remote system [3]. *Transfer Digital Credentials Securely (Work in Progress)* seeks to specify the transfer of digital credentials from one device to another. [3] Finally, *Key Transparency (Work in Progress)* seeks to specify tamper-evident storage of updates to public key infrastructure logging. [4] In alignment with its core mission, IETF specifications address a broad range of networked infrastructure without overemphasizing any single application or use case. In general, the standards maintained by IETF sit just below the technical stack layer discussed in this document.

- **W3C**: The W3C (World Wide Web Consortium) is the primary standards-development body for the Web. The relevance to this document is its coverage of higher-level schemas. In particular, *Decentralized Identifiers* specifies identifiers to real-world entities [15]. In addition, *Verifiable Credentials* specifies assertions about real-world entities [14]. Together, these two specifications underpin the larger technological cohesive and philosophically prescriptive ecosystem call SSI (self-sovereign identity). The SSI ecosystem seeks to provide individuals with control over their digital identity using technological innovations, most notably, cryptography and linked data. Our approach aligns with the high-level objectives of this ecosystem and attempts to improve on the scope, coherence, and rigor of existing frameworks.

- **ISO**: ISO (International Organizations for Standardization) establishes standards for many different aspects of technology and manufacturing from a distinctive government-centric perspective. The relevance to this document is its coverage of government use cases. In particular, *Mobile Driver's License* specifies digital driver's licenses for mobile devices [7]

---

[3] https://datatracker.ietf.org/wg/tigress/
[4] https://datatracker.ietf.org/wg/keytrans/about/

This standard introduces *mDocs* to extend the standard to other identifying information. In contrast to the other specifications mentioned in this section, ISO mobile driver's licenses are specifically scoped to a single use case (driver's licenses), stakeholder (governments), technical stack (a single mobile phone), and workflow (in-person presentation). We believe it is accurate to say that this standard is more focused on near-term adoption from traditional providers than it is on long-term support of advanced technology. For our purposes, mobile driver's licenses are one of many motivating use cases.

## 2.2. Decentralized Web

This section describes two categories of decentralized web technologies: DLT (distributed ledger technology) and decentralized social media. DLT is a class of technology that seeks to facilitate the secure sharing of state across distributed systems. The subset of DLTs that are most relevant to this document are those that seek to facilitate interoperability between different ledgers. Two exemplar DLTs with distinct approaches to interoperability are:

- **Polkadot**: implements a "layer 0" blockchain that coordinates application-specific client blockchains [19].

- **Cosmos**: implements an "Inter-Blockchain Communication protocol" that facilitates transactions between peer blockchains [9].

For our purposes, Polkadot and Cosmos are two useful data points in the general space of interoperable DLTs that we seek to abstractly describe. Meanwhile, decentralized social media is a growing industry concerned with the creation of social media platforms that are not controlled by any single organization. The following are examples of efforts to build decentralized social media:

- **Solid**: builds on Linked Data principles to facilitate general-purpose "pods" for users to interact with each other [13].

- **Mastodon**: builds on W3C ActivityHub protocol to create a Twitter-like service within the "fediverse" [18]

- **Bluesky**: builds on W3C Decentralized Identifiers and implements the custom AT Protocol to create a Twitter-like service [8]

The broad definition of digital identity that underpins this document includes social media interactions as a first-class example of identity information. Consequently, decentralized social media platforms are a key motivating use case in our technical discussion.

## 3. INFRASTRUCTURE

This section describes the technical foundations of the synchronic web. At this level of abstraction, the system is best understood as a method for sharing state in a peer-to-peer network. The components described in this section can underpin, but is not itself, a digital identity infrastructure.

## 3.1.    Design

Our model represents all information as binary hash trees, often called Merkle trees. Leaf nodes contain semantically meaningful expression while intermediate nodes contain cryptographic hashes of their child nodes. Figure 3-1 illustrates an example tree that encodes a simple mathematical expression. More interesting examples are the syntax tree of programming languages and the history of a state machine. Assuming a sufficiently collision-resistant hash function, there is a 1:1 relationship between the value of the root hash and the contents of the tree itself. Consequently, we can effectively refer to any member of the set of all computationally meaningful information using a fixed-size string.

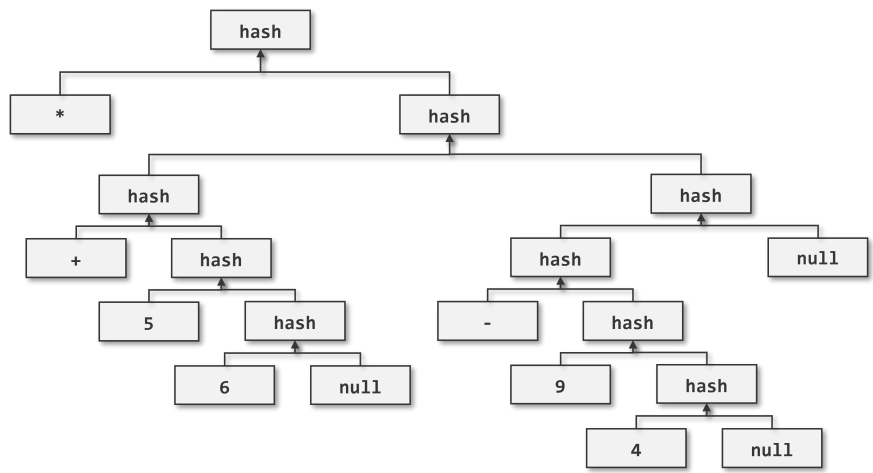

**Figure 3-1 Binary Hash Tree Example**

- **Entangled Nodes**: Merkles trees can themselves be the leaves of other Merkle trees. Consequently, the root of a tree could represent multiple semantically coherent expressions originating from multiple independent locations. Figure 3-2 illustrates an example of three related nodes that are each state machines. Since the Merkle tree in Node 3 contains the root of the Merkle trees of both Node 1 and 2, we say that the state of Node 1 and 2 are *entangled* with Node 3. Due to this setup, Node 1 and 2 can now perform joint computations in a way that makes use of the unique, secure context of their relationship to Node 3.

- **Entangled Networks**: The entanglement mechanism is transient. Consequently, the root of a tree could represent an arbitrary number of semantically coherent expressions originating from an arbitrary number of independent locations. Figure 3-3 provides a conceptual visual of an arbitrarily sized network of entangled states. In principle, the size of this network could approach and exceed the size of all connected devices in the modern Web, forming a connected graph of entangled states across the Internet. We hypothesize that such a network will someday emerge. This is the structure that we call the synchronic web.

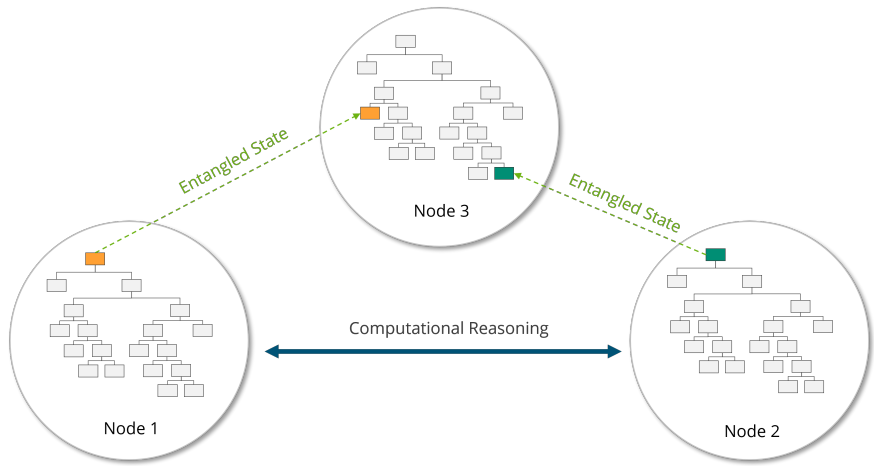

**Figure 3-2 Entangled Triad**

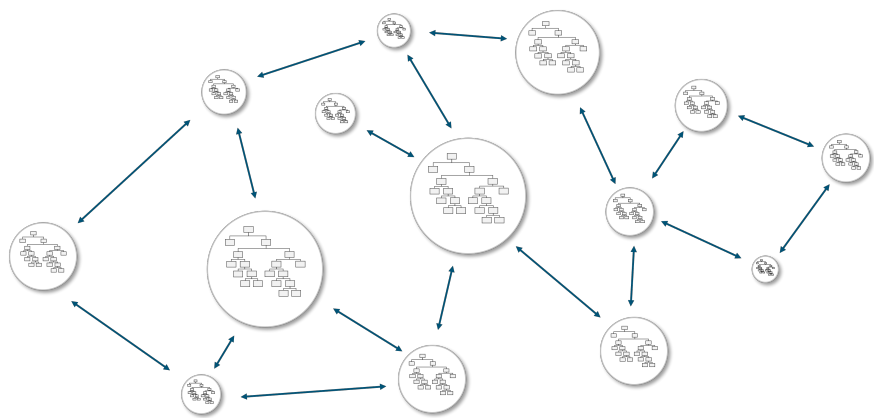

**Figure 3-3 Entangled Network**

## 3.2. Implementation

In practice, the situation described in 3.1 already exists. Today, the root of any non-trivial public blockchain likely includes, through an indirect sequence of cryptographically secure hashes, the root of millions of other semantically coherent trees (conservatively). However, we note a large emphasis on cryptocurrency use cases in the existing "Synchronic Web" and hypothesize that there exists a large space of opportunities to build toward alternative domains. To this end, we are actively developing software to explore these use cases. The most current piece of software that we maintain is an SDK (software development kit). Using this SDK, it is possible to build data structures that represent computable semantic expression. The primary language needed to use the SDK is a variant of Lisp [5]. Moving forward, we intend to support integrations with other programming languages and legacy services while preserving Lisp as the first-class language.

# 4. APPLICATION

This section describes our technical model of digital identity. Throughout the section, we will use terms from the SSI trust triangle—*issuer*, *holder*, and *verifier*—to refer to interacting parties. In addition, we will use the term *trust anchor* to refer to an issuer that the verifier depends on for ground truth.

## 4.1. Microstructures

In this section, we describe microstructures in the identity infrastructure. Microstructures model relationships between small numbers of nodes, when aggregated, can form larger macrostructures.

- **Verifiable Links**: A verifiable link is a periodic, directed, long-lived relationship where one node entangles its state with another and retains the evidence of entanglement in its state. Figure 4-1 provides an example. In this figure, an identity holder (dark blue) entangles its state with an issuer (light blue) for four complete rounds.
- **Verifiable Hubs** A verifiable hub is a complete set of all links that originate at the same node. Figure 4-2 provides an example. In this figure, the holder (dark blue) has links to five other issuer nodes (light blue) for four complete rounds. The salient property of hubs is completeness; a verifier can only verify a hub if the prover can formally define the set of relevant links for the relevant point in time in the relevant context.
- **Verifiable Chains**: A verifiable chain is a series of verifiable links that connect two nodes across one or more intermediary nodes. Figure 4-1 provides an example. In this figure, a verifiable chain connects the holder (dark blue) to the trust anchor (green) through three intermediary nodes (light blue) for two complete rounds. The salient property of chains is connectedness; a verifier can only verify a chain if it can trace information from the holder to the trust anchor and back. Consequently, the length of a chain determines the latency between the time that a state can be read and the time that it can be verified.

---

[5]Currently, a modified fork of s7 scheme (https://ccrma.stanford.edu/software/s7/s7.html)

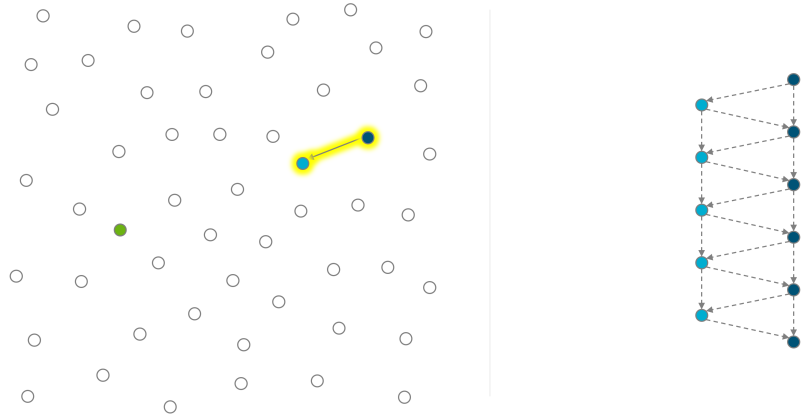

**Figure 4-1 Persistent Link**

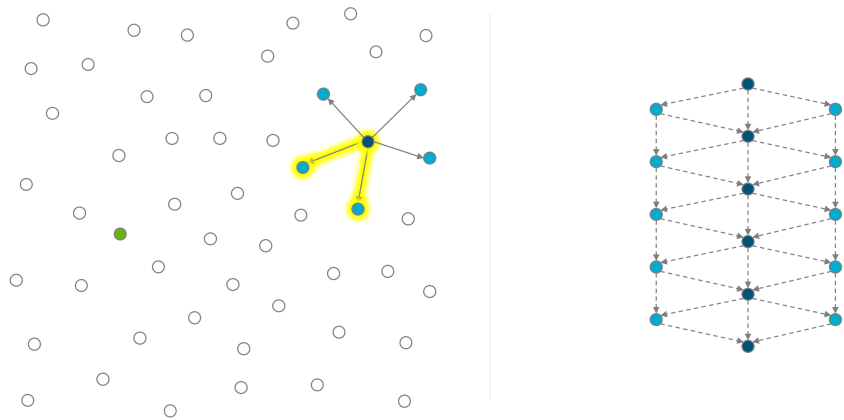

**Figure 4-2 Persistent Hub**

## 4.2. Macrostructures

In this section, we describe macrostructures in the identity infrastructure. Macrostructures model network topologies that can be organically formed from smaller microstructures. The shape of these structures depend on two competing set of factors. On one side, the theoretical benefits of *completeness* and *connectedness* creates an upward pressure on the number of links. On the other side, the practical costs of storage, bandwidth, and computational creates a downward pressure on the number of links. In this context, macrostructures describe different points in the trade-off space that are plausibly optimal for different use cases. We hypothesize that the synchronic web will continually evolve into a coherent composition of locally optimal macrostructures. [6] The following sections discuss the four structures provided in Figure 4-4. For each structure, we provide a description of the topology, examples of applicable use cases, and commentary on the optimizations enabled in a synchronic web (vs non-synchronic-web) paradigm.

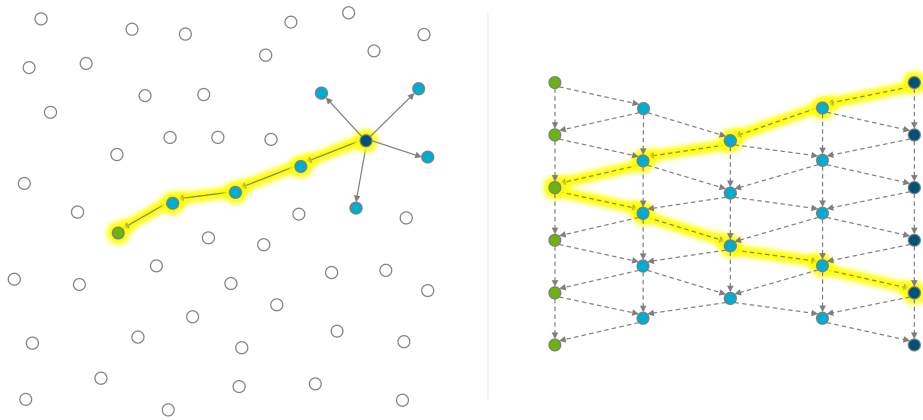

**Figure 4-3 Persistent Chain**

- **Centralized**: In a centralized structure, many identity holders entangle their state with a single trust anchor. The topology of this structure mirrors many simple identity information systems that exists today, for example: electronic passports issued by a single government agency, employee identities controlled by a single enterprise directory server, and cryptocurrency wallets operating on a single-layer blockchain. Notably, a centralized topology can still be useful even if the trust anchor does not store any of the holders' actual data. To improve scalability and confidentiality, it only needs to store a single hash representing the state of the network at each step in order to maintain a secure identity ecosystem for the holders.

- **Federated**: In a federated structure, identity holders entangle their state with a chance of intermediaries that all point back to a single root federation authority. The topology of this structure mirrors many large-scale identity information systems that exist today, for example: Real ID driver's licenses standardized by the Federal government and issued by individual states, Domain Name System issued by individual servers and ICANN, and cryptocurrency wallets operating on a multi-layer blockchain In a federated topology, the utility of sharing synchronic web states is to mitigate the risk of intermediary misbehavior. As long as holders track (potentially through a gossip protocol) the same root hash for the federation authority, they are protected from maliciously inconsistent intermediary issuers.

- **Interoperated**: In an interoperated structure, identity holders entangle their state with more than one trust anchor. The topology of this structure mirrors many emerging efforts to implement interoperable Web identities, for example: individuals maintaining dual citizenship with multiple countries, OpenID connect replying parties that support multiple issuers [12], and cryptocurrency wallets that support multiple blockchains. While semantic interoperability between issuers is the primary challenge in this setup, the ability to secure share state through the synchronic web enables a more secure way to facilitate interoperability.

---

[6]We hypothesize that the resulting network will be a case of the "Small World" phenomenon [17].

- **Decentralized**: In a decentralized structure, nodes entangle their state with each other in peer-to-peer relationships. The topology of this structure mirrors the highest level of many consequential networks in the real world, for example: the relationship between countries in international relations, the relationship between Autonomous Systems in Internet routing, and the interoperability between disparate blockchains in cryptocurrency networks. In a decentralized topology, the network of synchronic web nodes effectively participates in a consensus protocol. The nature of the coordination determines the nature of guarantees that can be made about the resulting network state.

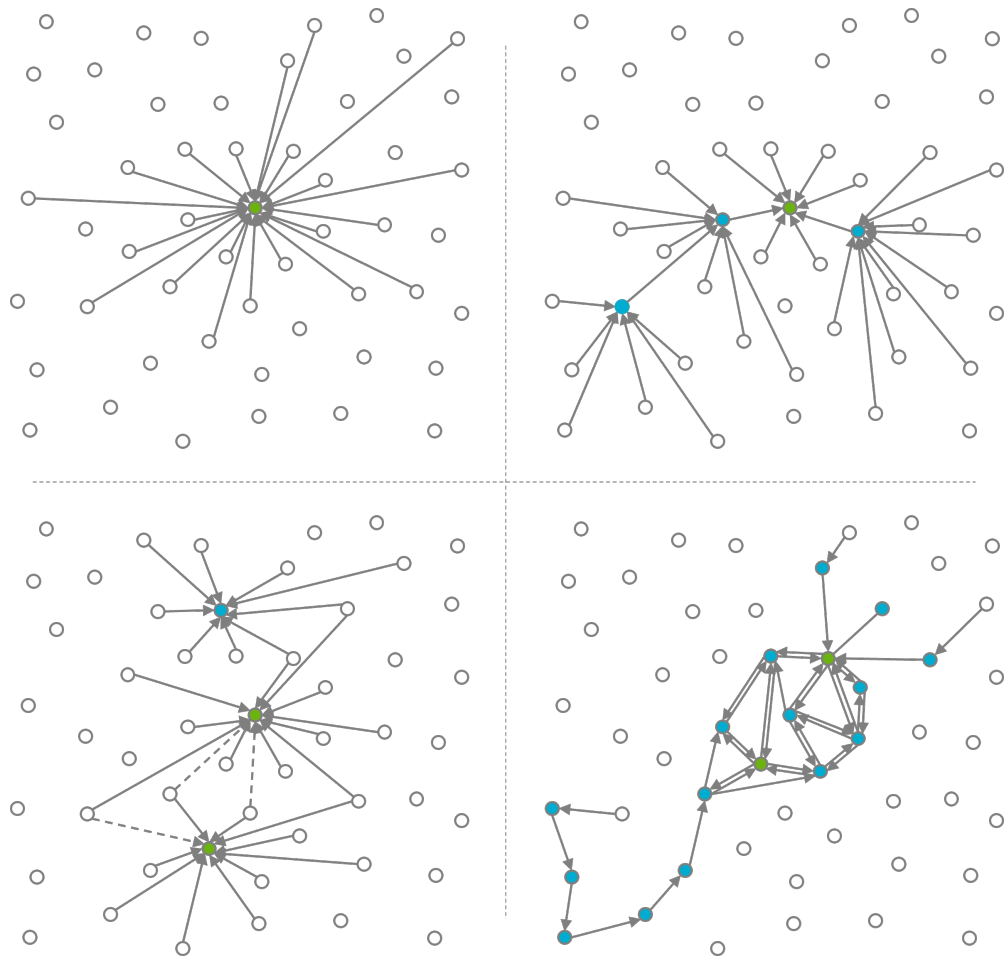

**Figure 4-4 Canonical Topologies**

# 5. DISCUSSION

This section discusses the technical model described in Section 4. We begin by enumerating special topics that cover key aspects of the current model and end by speculating on future research requirements and opportunities.

## 5.1. Operations

The special topics discussion in this section aims to meet some combination of the following two objectives: first, to consolidate our definition of digital identity with more conventional intuitions, and second, to explore the current limits of our model. Each topic could plausibly motivate a standalone research project.

- **Physical Binding**: Physical binding is the process in which a verifier connects a digital identity to a real-world entity. This process involves a wide range of technology related to software, hardware, and "wetware". One effort attempting to standardize physical binding at Internet scale is the Web Authentication protocol [16]. In general, the physical binding process is orthogonal to the topics in this document. However, one topic that might affect both domains is the trustworthiness of the entities involved in the binding process. For example, synchronic web notions of digital identity may be useful for establishing the reputation of a specific biometric device, the organization that manufactures the device, or the government agency that uses the device.

- **Credential Revocation**: Credential revocation is the process in which an authorized party invalidates a previously issued credential. Revocation is naturally supported on the synchronic web. Notably, all credential revocation systems exist somewhere in the trade-off space between integrity and confidentiality. For instance, on one end of the spectrum, the Online Certificate Status Protocol facilitates near-instantaneous revocations but leaks information about credential usage to the issuer [10]. On the other end of the spectrum, the visual inspection of a physical driver's license does not leak any information but does not check for revocations at all. In the synchronic web, we model the former as a credential that is controlled by the issuer and the latter as a credential that is controlled by the holder.

- **Access Control**: Access control is the framework that allows authorized parties to restrict unauthorized parties from reading information or writing state. In the current Web, access control is somewhat orthogonal to digital identity. This is because authentication deals with the stochastic, asynchronous task of message-passing (credentials) while authorization deals with the deterministic, synchronous task of rule-writing (access control policies). Examples of open frameworks for creating access control policies are XACML [1] and NGAC [5]. One way to describe the utility of the synchronic web is to abstract away the stochasticity and asynchrony of distributed systems. At scale, we hypothesize the ability to more cleanly unify authentication frameworks with authorization frameworks to create more natural access control policies for the web.

- **Key Recovery**: Key recovery is the process that allows the holder to regain control of its identity after losing the controlling key. Beyond immediate measures like separating short-lived keys from long-lived keys, recovery requires, at some point, trust in one or more external parties who are authorized to perform the recovery. For example, the social security administration has the authority to "recover" accounts by replacing the social security number. More generally, many forward-looking identity systems aspire to build "social recovery" systems in which some majority threshold of the holder's social circle is authorized to perform the recovery. The synchronic web naturally supports the full spectrum of social recovery mechanisms through the concept of verifiable hubs. Provided $n$ entangled issuers, it is possible to define explicit semantics declaring that $m$ of them are authorized to help the holder establish a new secret key in a recovery situation.

## 5.2. Costs

In this document, we have discussed many reasons why individuals and organizations may want to embrace a future digital identity paradigm rooted in the synchronic web. Although there many reasons they may not want to do so, the reasons are primarily practical in nature: time, effort, and money. However, there is one unavoidable theoretical cost: the loss of plausible deniability. Plausible deniability, more formally called repudiation, is the ability of a party to deny their authorship or ownership of some piece of information. There are many valid situations in which an individual may wish to assert the ownership of a piece of information to some verifiers, but not others. For instance, a whistle-blower communicating on an encrypted network may wish to authenticate themselves to a trusted journalist, but not to the public. In another example, an individual in the European Union might want to post some social media message today but may want to exercise their "right to be forgotten" under the General Data Protection Regulation tomorrow. Much like the relationship between reputation and privacy, plausible deniability is diametrically opposed to the core data integrity purpose of the synchronic web. However, if no system exists that can guarantee both perfect integrity and perfect deniability, then perhaps a system exists that can navigate the trade-off space between the two desired properties.

More formally, we wish to create a system such that Peggy, as a prover, can convince Victor, as a verifier, that some assertion on her content is (1) correct according to the data committed on the synchronic web (2) is derived from data at a particular location committed in the Synchronic web, and at a particular location in the committed Merkle tree of data (3) design the proof and protocol such that though Victor is convinced of the veracity and integrity of the proof, it will not be possible for Victor to take the proof and replay it to convince some third party that Peggy was the prover and the assertions hold with respect to her. Currently, some primitives exist to be able to do these things individually, in specialized settings, or with non-standard assumptions. The two chief cryptographic techniques conjectured to be most useful in addressing this are ZKPs (zero-knowledge proof) and MAC (message authentication code). The most significant problem anticipated is related to the key-agreement its assumptions/properties related to its use in the protocol, ZKP, and MAC. We seek to bring together a whole system binding the primitives together in way which maintains their security and still achieves our goal. The feasibility of designing such a system is an open research question.

## 5.3. Path Forward

This section describes the research path forward for both completing this document and future work that may extend or supersede it. First, the combined topics of cryptographically verifiable state and digital identity, by our definition, are incredibly broad. In time, we hope to continually refine both the messaging of the value proposition and the proposed technologies that these ideas enable. The next step that will help to alleviate this problem is the continued development of our proof-of-concept path-finder system. As we augment the existing prototype with increasingly sophisticated identity capabilities, we will be able to discuss concrete results rather than abstract possibilities. The two primary technical next steps are:

- **System Emulation**: A critical research path will be the emulation and analysis of the system at scale. Once a sufficiently networked prototype exists, we anticipate a compelling research opportunity to conduct distributed, agent-based experiments to understand the dynamics of a synchronic web digital identity system at large scales. Well-designed experiments will shed light on the practical limits of the value proposition and potentially uncover unexpected emergent behaviors. Sufficiently positive results would contribute to the case for real-world deployment.

- **Formal Methods**: To further capitalize on value proposition of a deterministic and synchronous digital identity framework, theoretical opportunities exist to provably qualify the space of security guarantees afforded by the synchronic web. For instance, languages like Rocq [2] and Idris [4] are promising tools for adding theoretical rigor to the existing synchronic web prototype.

Finally, all large-scale work on the World Wide Web is, at some level, rooted in open standards. In addition to required work to standardize the synchronic web itself, the framework that we have described in this document is, in many ways, ideal for propagating the notion of standardization itself. Today, the standards that underpin the world's largest information system depend on implicit reputations and informal compliance. In contrast, we can imagine a future paradigm in which the identities and semantics that drive this work are propagated as explicit branches in a global cryptographic data structure. The opportunity to not only build upon, but actively drive, such progress is the preeminent objective of our work.